\documentclass[aps,twocolumn,prd,groupedaddress,showpacs,showkeys]{revtex4}
\usepackage{graphicx}
\usepackage{subfigure}
\usepackage{dcolumn}
\usepackage{mathrsfs}
\usepackage{amsmath}
\usepackage{latexsym}
\usepackage{amssymb}

\begin{document}

\begin{minipage}[b]{2\linewidth}
\hspace{13.55cm}{Imperial/TP/07/TK/01}
\end{minipage}\hfill
\vspace{0.5cm}

\title{Kinematic Constraints on Formation of Bound 
States of \\ Cosmic Strings  -- Field Theoretical Approach}

\author{P. Salmi$^{1}$}
\email{salmi@lorentz.leidenuniv.nl}
\author{A. Ach\'ucarro$^{1,2}$}
\email{achucar@lorentz.leidenuniv.nl}
\author{E. J. Copeland$^{3}$}
\email{ed.copeland@nottingham.ac.uk}
\author{T. W. B. Kibble$^{4}$}
\email{t.kibble@imperial.ac.uk}
\author{R. de Putter$^{1,5}$}
\email{rdeputter@berkeley.edu}
\author{D. A. Steer$^{6}$}
\email{steer@apc.univ-paris7.fr}
\affiliation{$^{1}$Lorentz Institute of Theoretical Physics, 
University of Leiden, The Netherlands \\
$^{2}$Department of Theoretical Physics, University of the Basque
Country UPV-EHU, 48080 Bilbao, Spain \\
$^{3}$School of Physics and Astronomy, University of Nottingham, 
University Park, Nottingham NG7 2RD, United Kingdom \\
$^{4}$Blackett Laboratory, Imperial College, London SW7 2AZ, United Kingdom \\
$^{5}$University of California, Berkeley, CA 94720, USA \\
$^{6}$APC, Universite de Paris 7, Batiment Condorcet, 10 Rue Alice Domon et
L\'eonie Duquet, 75205 Paris, France}

\date{\today}

\begin{abstract}
Superstring theory predicts the potential formation of string networks 
with bound states ending in junctions. 
Kinematic constraints for junction formation have been 
derived within the Nambu-Goto thin string approximation. 
Here we test these constraints numerically in the framework of the
Abelian-Higgs model in the Type-I regime and report on good agreement 
with the analytical predictions.
We also demonstrate that strings can effectively pass through each 
other when they meet at speeds slightly above the critical velocity
permitting bound state formation. 
This is due to reconnection effects that are beyond the scope of the 
Nambu-Goto approximation.

\end{abstract}

\keywords{cosmic strings, string junctions, Abelian-Higgs model}
\pacs{11.15.-q, 11.27.+d, 98.80.Cq}

\maketitle

\section{Introduction}

There has been a revival of interest in cosmic strings since it was 
realised that many fundamental string theory models predict 
so-called {\it cosmic superstrings}.
The leap from string scale to cosmic dimensions is highly non-trivial.
At first sight there are many severe problems related to the formation 
and growth of superstrings to macrosopic scales, such as too high string 
tension and instability towards breaking up, as pointed 
out already in~\cite{Witten:1985fp}. 
Subsequent advances in fundamental string theory have 
changed this picture. 
All these issues can be circumvented and are naturally evaded in many models 
that possess otherwise desirable aspects from the point of view of providing 
successful phenomenology.
For instance, in the scenarios based on warped compactifications 
the warp-factor can reduce the string tension. Furthermore, in models of brane 
inflation not only are cosmic strings produced at the end of 
the inflationary epoch~\cite{Sarangi:2002yt,Majumdar:2002hy}, 
but it has been argued in great abundance too~\cite{Barnaby:2004dz}.

Considerations inspired by superstring theory suggest
a particular kind of cosmic string network 
that consists of fundamental F-strings, Dirichlet D-strings 
(or more precisely D1-branes) and {\it bound states} of these two, 
known as $(p,q)$-strings as an abbreviation for $p$~F-strings 
and $q$~D-strings~\cite{Copeland:2003bj}. 
The presence of $(p,q)$-strings brings an additional feature to these 
networks compared to those made of only one type of solitonic strings. 
Namely, where two different types of string meet at a point and form a bound 
state leading away from that point, there is a {\it junction} in the network.
Another property of cosmic superstrings is that their reconnection 
probability $P$ can be small, $P \ll 1$ 
(for a review see~\cite{Polchinski:2004ia} and references therein).

A substantial body of work has dealt with modelling of cosmic string networks, 
which is a challenging task due to the combination of several scales involved 
and the non-linear nature of the problem.
For instance, properties of networks at small scales are still not 
entirely understood and have been under intense analytic study 
recently~\cite{Polchinski:2006ee,Polchinski:2007rg,Dubath:2007mf}.
However, it is well-established that string networks lose energy efficiently. 
A scaling string network contributes only a fixed, tiny fraction 
of the total energy budget of the Universe, a
property that has made cosmic strings viable, 
in contrast to some other defect models, and generated the appeal of 
cosmic strings from the very start~\cite{Kibble:1976sj}.

For the above mentioned reason the effect of junctions and bound states 
on the evolution of the networks is of 
profound interest: instead of reaching a scaling regime 
the network could end in a frozen-out state and start to dominate the energy 
density of the Universe, which cannot be tolerated in the standard 
cosmologies. 
To date, the evolution of networks with junctions has already 
been studied in several models.
Good evidence for scaling was reported in~\cite{Hindmarsh:2006qn} with an 
SU(2)/${\mathbb Z}_{3}$ model of global strings, the junctions being 
global monopoles. Another field theory study~\cite{Rajantie:2007hp} 
used a model involving two sets of U(1) gauge fields 
(see also~\cite{Saffin:2005cs}). There bound states are reported to have a 
significant effect on the network in the absence of long-range interactions, 
whereas in the case of global strings junctions play a minor r\^ole.
Other studies 
include~\cite{Copeland:2005cy,Leblond:2007tf,Avgoustidis:2007aa}, 
which concluded that, even being conservative, the networks can scale 
also in the presence of junctions.

If the bound states and junctions are of major importance, then 
a natural question to pose is under which conditions they form.
This was examined analytically based on the Nambu-Goto action 
in~\cite{Copeland:2006eh,Copeland:2006if}. It was shown that 
strong kinematic constraints apply to the formation of the bound state.
The purpose of this study is to test these constraints in a 
field theory set-up.
We work within the Abelian-Higgs model in the type-I regime, 
investigating when 
the intersection of strings results in the formation of a bound state 
(also called a {\it zipper}). 
This topic has been addressed already before both 
analytically~\cite{Bettencourt:1994kc} and with numerical 
experiments~\cite{Bettencourt:1996qe}. Here we revisit it due to 
the renewed interest in more complicated networks 
with a spectrum of string tensions
and with improved computational resources available.
Before presenting the results, we briefly review the outcome 
from the Nambu-Goto 
approximation and introduce the model together with the 
numerical approach.

\section{String Junctions}

Consider a straight string making an angle $\alpha$ with the 
positive $x$-axis and another one with an angle $-\alpha$ 
(the total angle between the strings being thus $2 \alpha$). 
Both strings are on the $xy$-plane and have a velocity $v$ along the $z$-axis 
with opposite directions 
and have string tensions $\mu_1$ and $\mu_2$, respectively.
Once they intersect, these can potentially form a third string, 
an ``$x$-link'' which has tension $\mu_3$ 
(we follow here the notation introduced in~\cite{Copeland:2006if}; 
if $\mu_1=\mu_2$ then the $x$-link is indeed positioned along the $x$-axis).

The approach in~\cite{Copeland:2006eh,Copeland:2006if} is based on 
studying the action at a junction where three strings meet. 
The kinematic constraints follow from the requirement that the 
total length of the progeny string must increase.
The allowed parameter region for the link formation can be determined 
at least numerically for any combination of string 
tensions~\cite{Copeland:2006if}. 
However, when the colliding strings have the same tension, 
denoted hereafter by $\mu_1$, it is possible to express 
this requirement in a simple closed form
\begin{equation}
\alpha < \arccos\big(\frac{\gamma \mu_3 }{2 \mu_1}\big) \, , 
\label{inequality}
\end{equation}
where $\gamma = 1/\sqrt{1-v^2}$ and
necessarily $\mu_3 < 2 \mu_1$. 

Here we want to test this result numerically 
in a field theory where strings are solitonic objects 
with a {\it non-zero~width} and see if bound state formation is 
the dynamically preferred process.

\section{Model and Numerical Implementation}

The Abelian-Higgs model is governed by the Lagrangean 
\begin{equation}
{\mathcal{L}} =
({\partial}_{\mu} + i q A_{\mu} ) \, \phi \, 
({\partial}^{\mu} - i q A^{\mu} ) \, \phi^{\dagger}
- \frac{1}{4} F_{\mu \nu} F^{\mu \nu}   - 
\frac{\lambda}{4} (|\phi|^2-\eta^2)^2 \, .
\nonumber
\label{abelianhiggs}
\end{equation}
The model has vortex solutions~\cite{Nielsen:1973cs}, in which 
the scalar field can be expressed, with the help of a function $f$ 
of the radial distance $r$ only, as follows:
\begin{equation}
\phi = f(r) \, e^{i n \theta},
\nonumber
\label{nielsen-olesen}
\end{equation}
where $n$ is the winding number.
For $n=1$ the vortex solutions are topologically stable 
for any value of the scaled 
coupling, $\beta = \lambda/2q^2 = m_{\rm scalar}^2 / m_{\rm gauge}^2$, 
where $q$ is the gauge coupling, $m_{\rm scalar} = \sqrt{\lambda}\eta$ 
and $m_{\rm gauge} = \sqrt{2} g \eta$. 
The domain $\beta < 1$ is known as the type-I regime. 
There a string with winding number $n$ has lower energy 
than $n$ strings with winding number $n=1$ together.
In particular, two $n=1$ strings can merge to form 
an $n=2$ string~\cite{Jacobs:1978ch}.
In other words, using the previous notation, $\mu_3 < 2 \mu_1$, and 
thus this regime provides a testing ground for the formation of links.

The numerical code used to evolve two boosted strings on a lattice 
has been reported in~\cite{Achucarro:2006es}. 
The boundary condition employed here was introduced in~\cite{Matzner:1988}:
the fields on the boundaries are updated as though the strings moved 
undisturbed at constant, initial velocities.  
Of course, this means that the simulations cannot be trusted much 
beyond the time when the kinks on the strings generated by their 
interaction reach the boundary.
The majority of the simulations were carried out in a computational 
grid of size $400^3$, setting the lattice spacing $dx$ to be $0.2$ 
(the physical size of the lattice in linear dimension being therefore $80$) 
and the time step $dt=0.065$ 
($dx$ and $dt$ here in units of $(q \,\eta\,)^{-1}$). 
The robustness of the results has been 
tested by varying the size of the simulation box, lattice spacing and 
the initial separation of strings. All the snapshots 
presented in the following section are from simulations performed 
on a $600^3$ grid, setting $dx=0.15$.

\section{Results}

\begin{figure*}
\begin{center}

\includegraphics*[width=.4\textwidth]{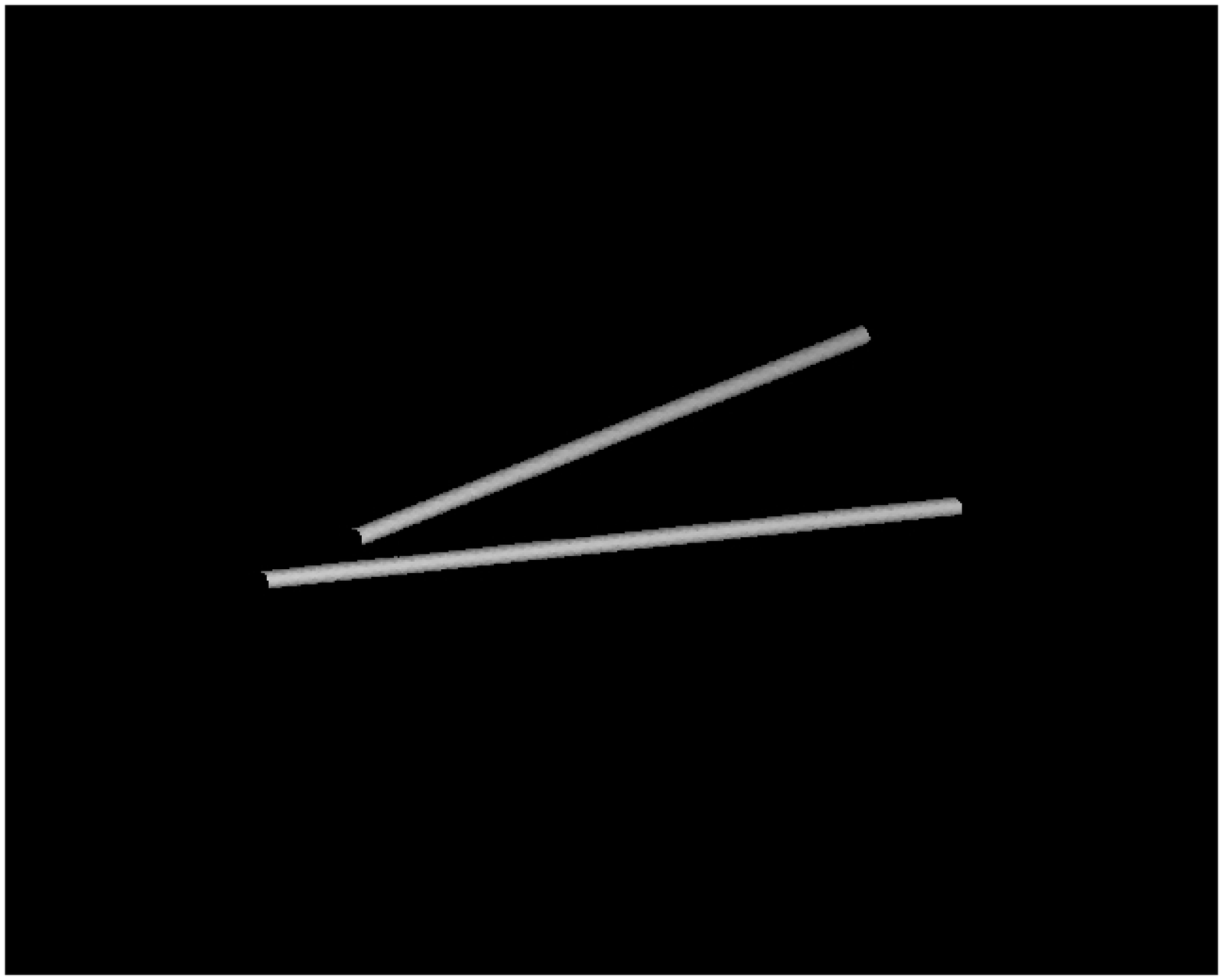}
\includegraphics*[width=.4\textwidth]{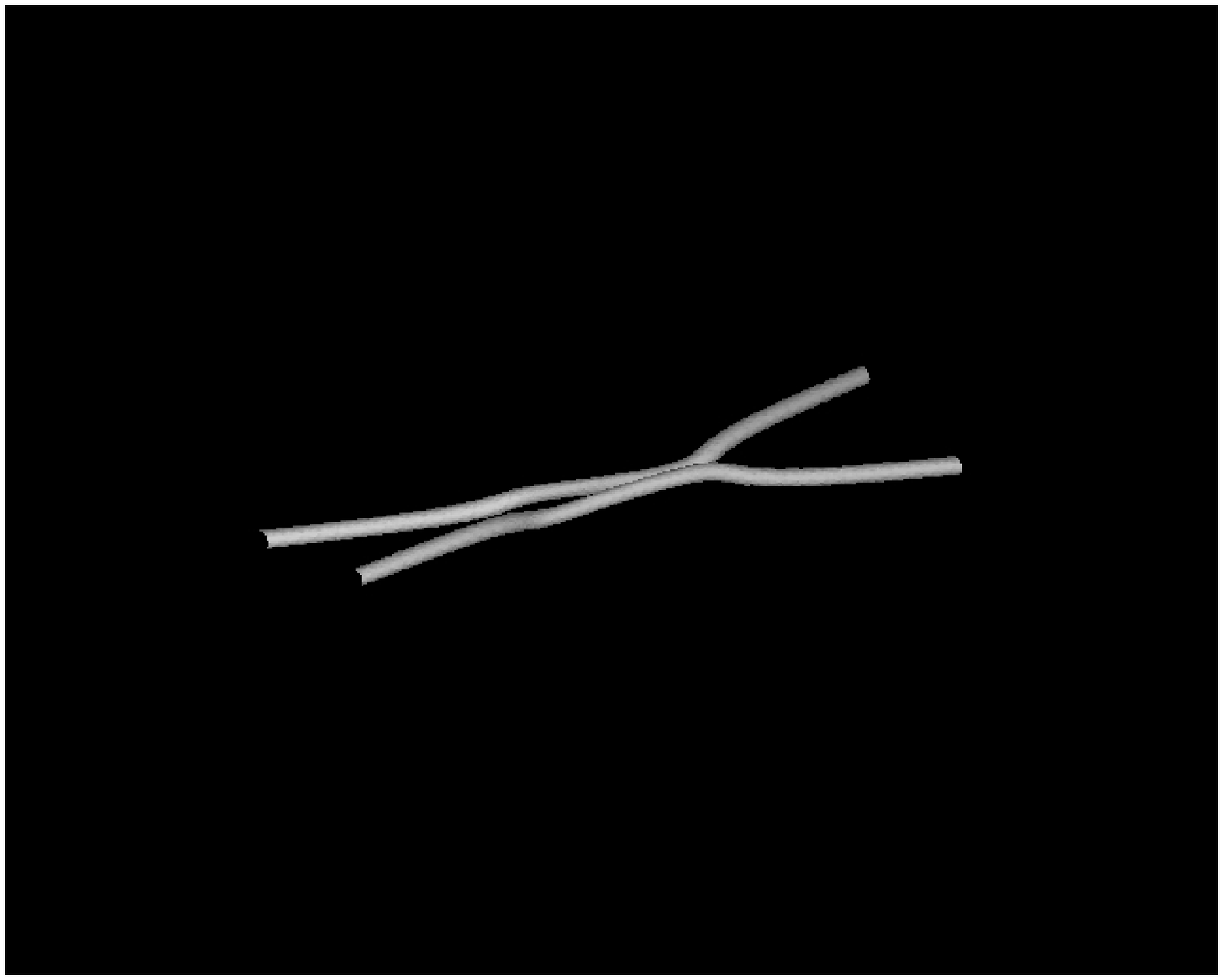}
\includegraphics*[width=.4\textwidth]{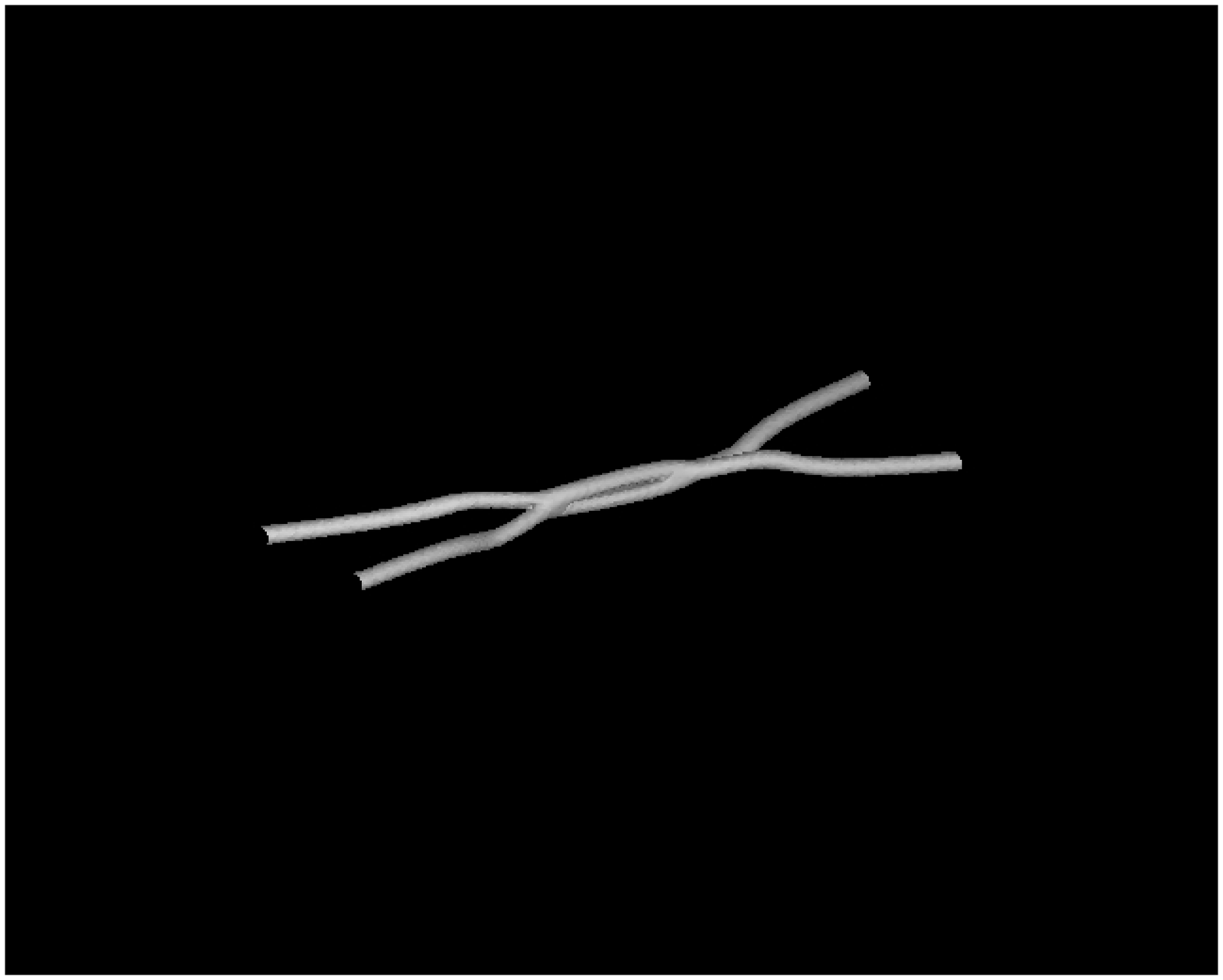}
\includegraphics*[width=.4\textwidth]{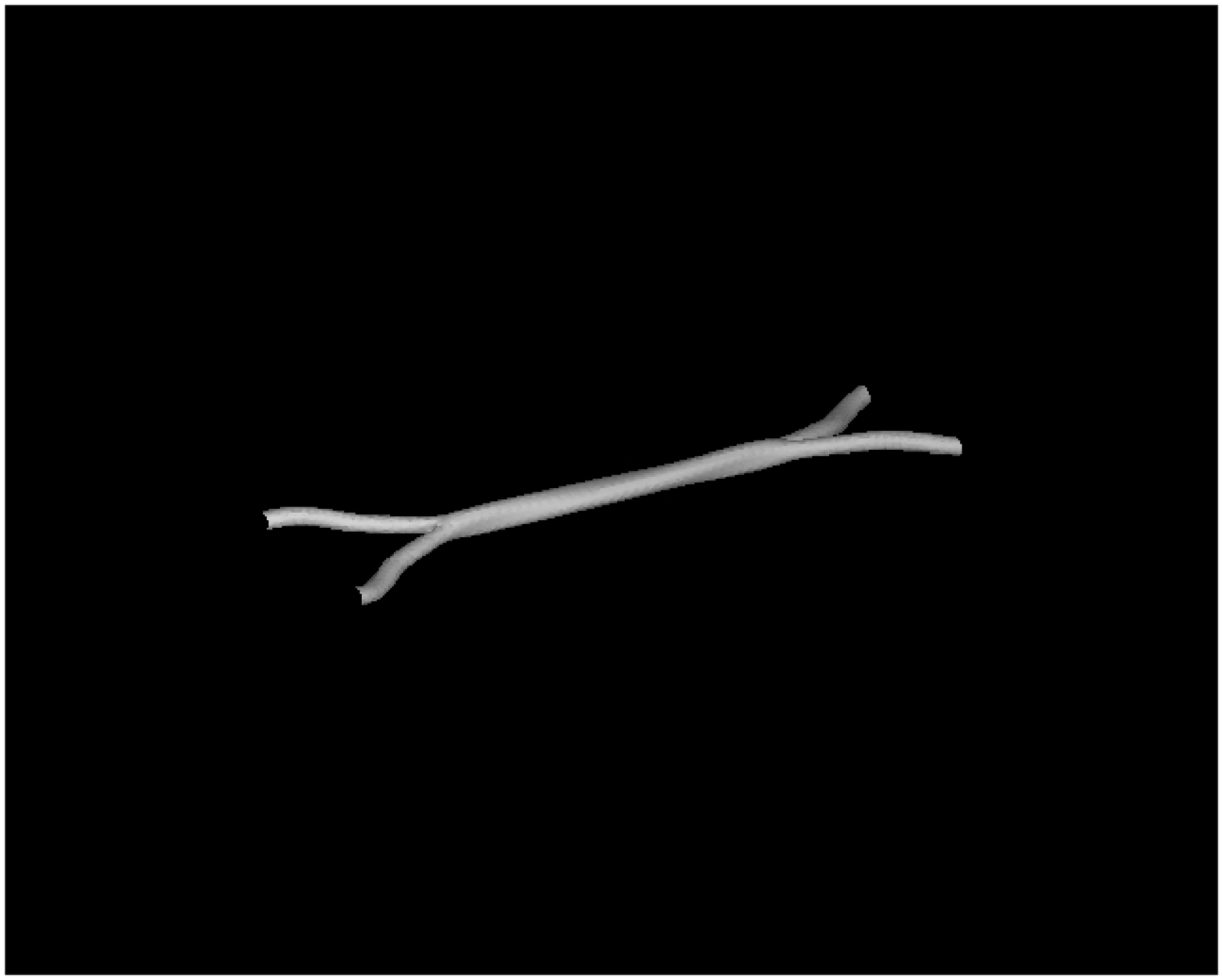}

\vspace{1.0cm}

\caption{\label{f:merge} Snapshots (from left to right and top to bottom) 
showing constant energy 
density isosurfaces in a simulation of two strings with $n=1$ 
at $\beta=0.125$ colliding with velocity $v=0.1$ 
(top one moving downwards and vice versa), at 
an angle $\alpha=15^{\, \circ}$, and forming an $x$-link as 
indicated by the $\circ$ in Fig.~\ref{f:constraints}.}

\end{center}
\end{figure*}


\begin{figure*}
\centering

\includegraphics[width=.45\textwidth]{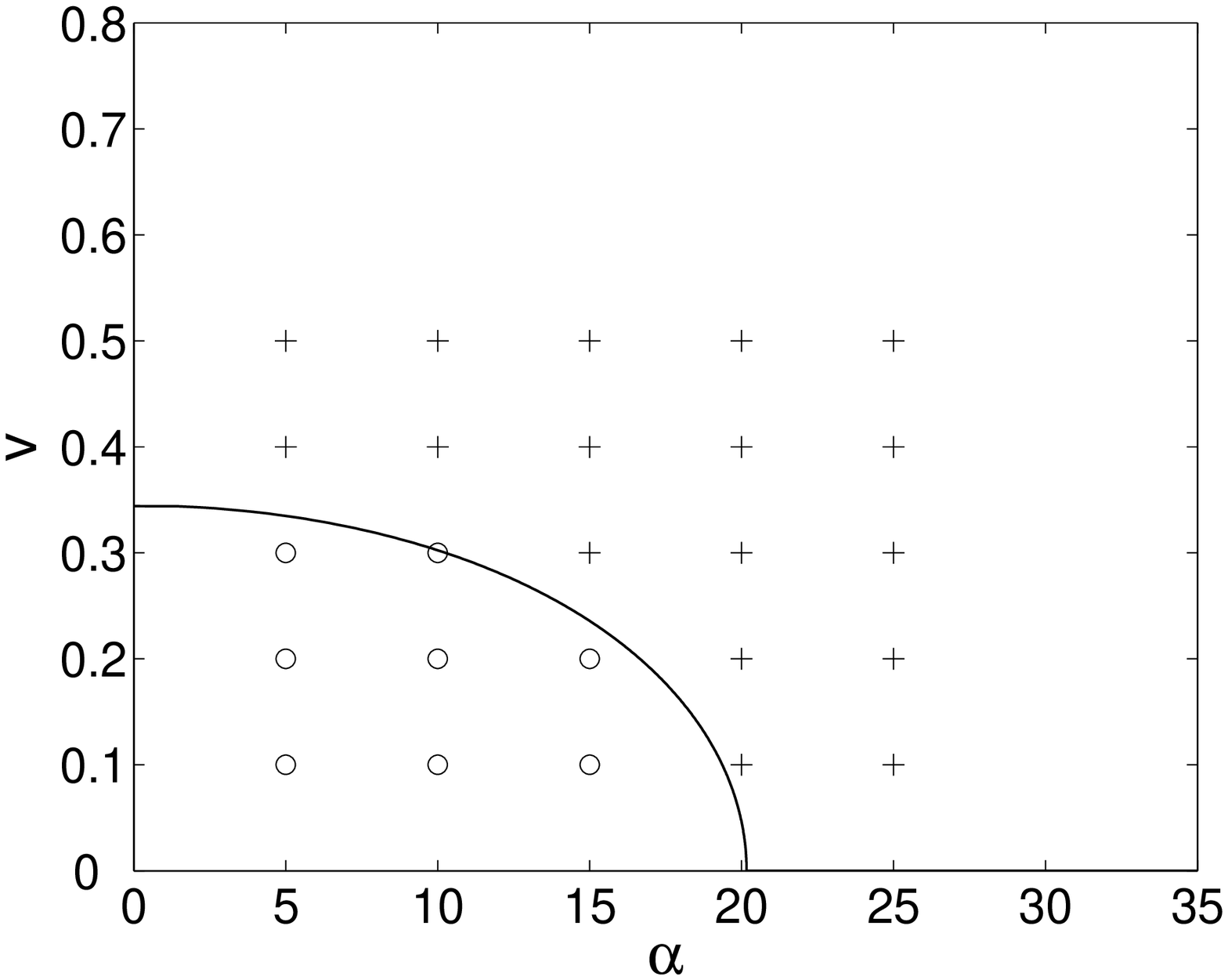}
\includegraphics[width=.45\textwidth]{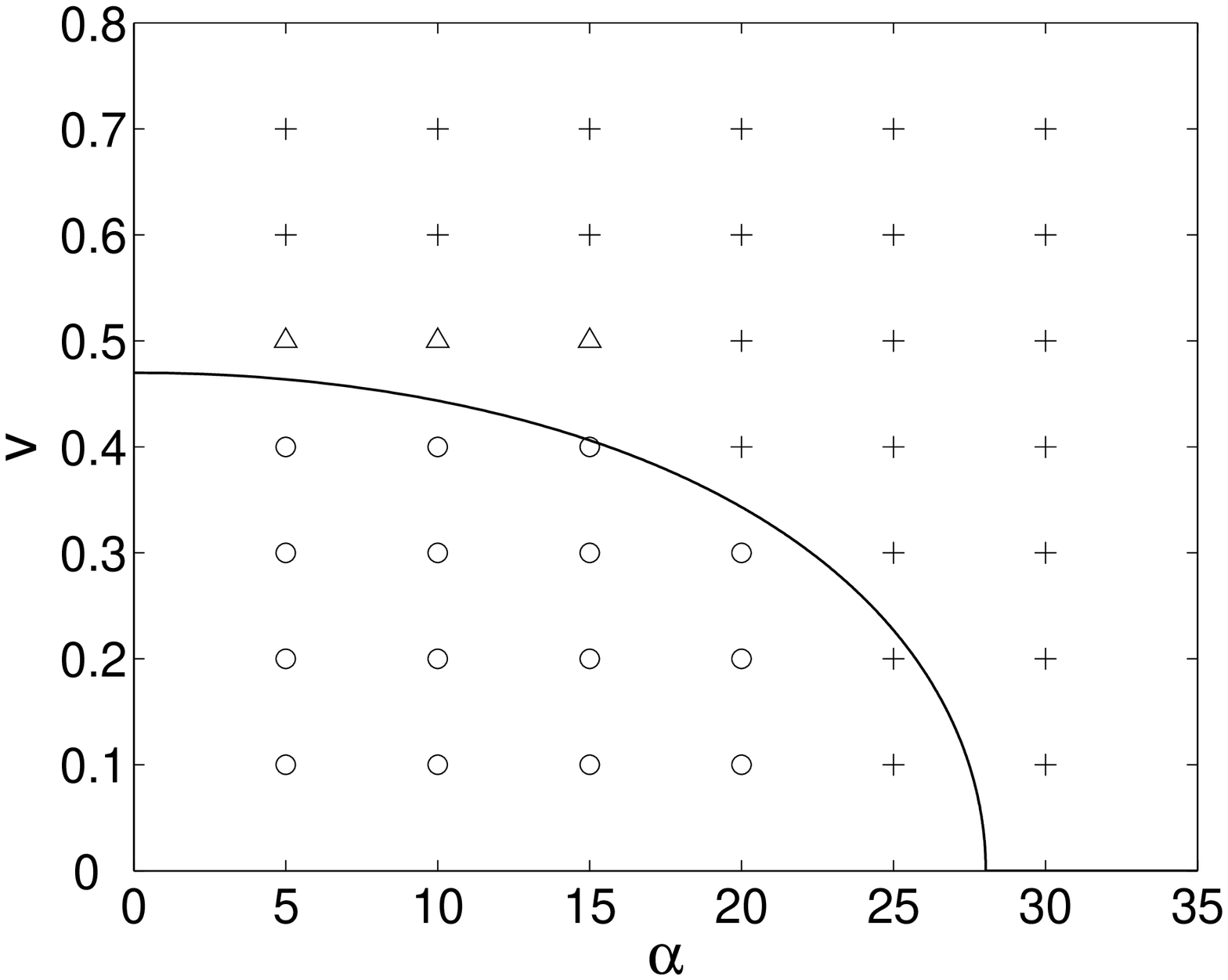}

\vspace{1.0cm}

\caption{\label{f:constraints} Kinematic constraints in the $(\alpha,v)$-plane 
for $\beta=0.36$ (left) and $\beta=0.125$ (right).
The solid line shows the curve of equality in~(\ref{inequality}), 
+ marks the events of single intercommutation, whereas $\circ$ indicates 
strings merging together to form an $x$-link, a string with 
winding number $n=2$. Events where strings pass through each other by 
intercommuting twice are shown by $\vartriangle$.}

\end{figure*}


\begin{figure*}
\centering

\includegraphics*[width=.4\textwidth]{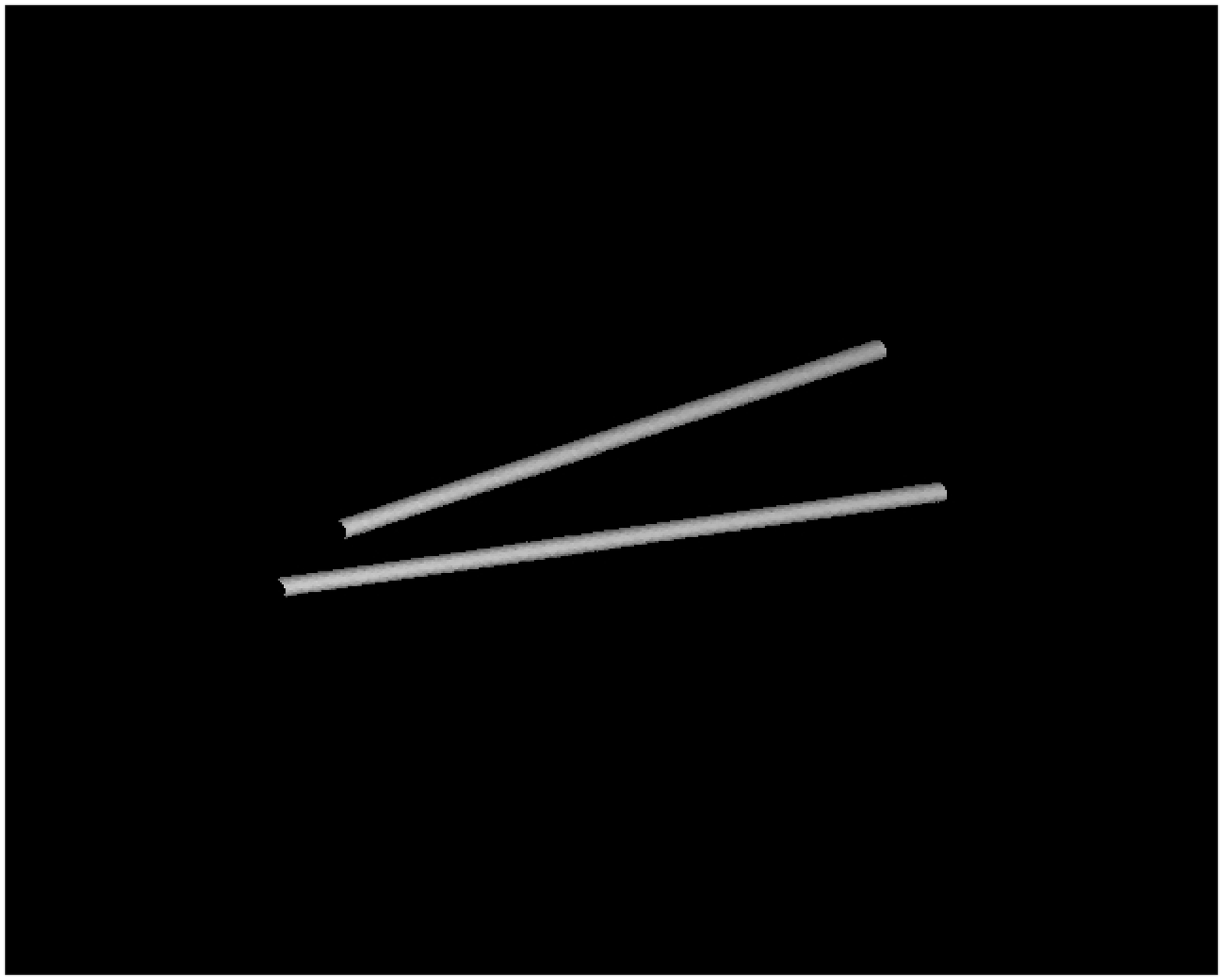}
\includegraphics*[width=.4\textwidth]{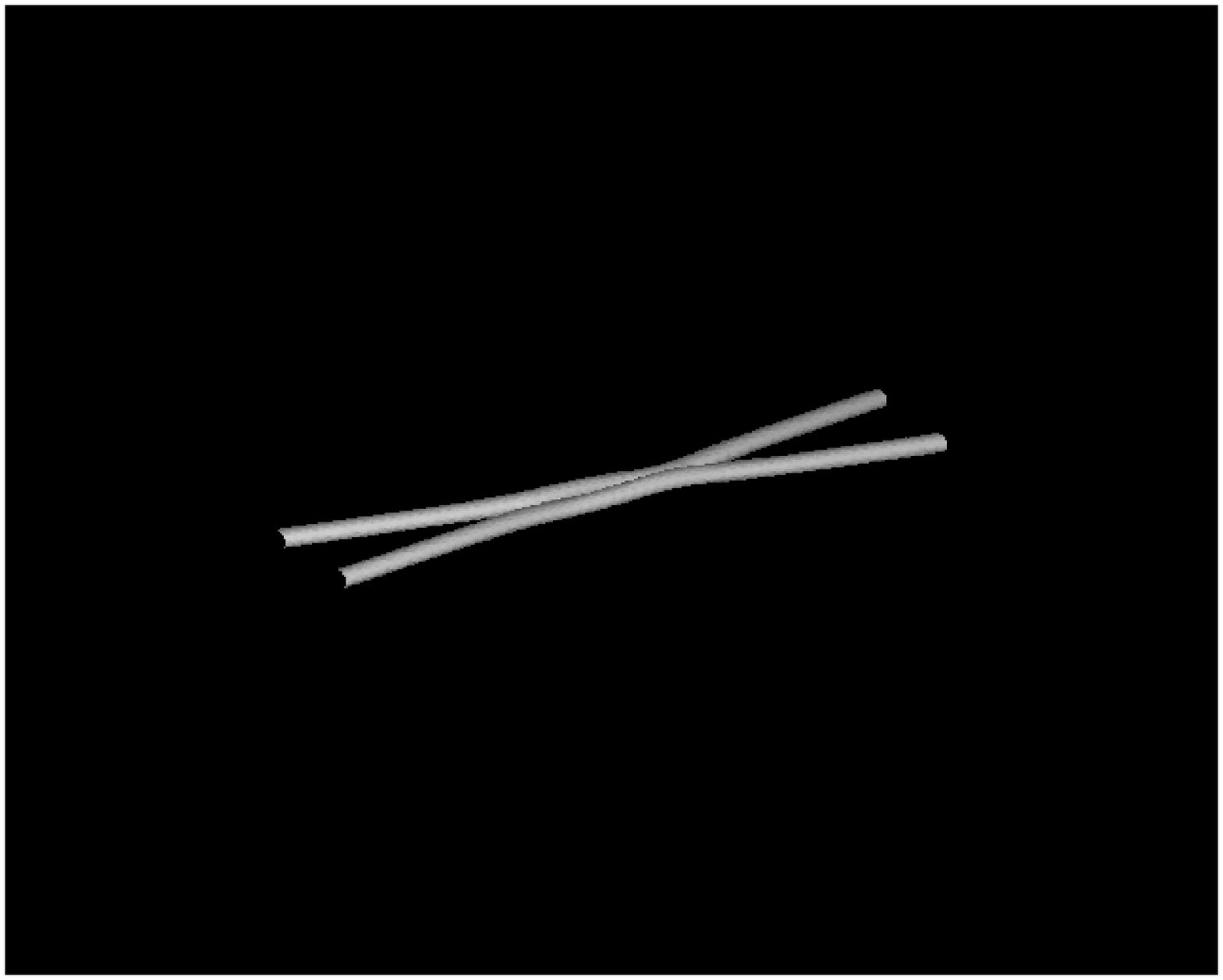}
\includegraphics*[width=.4\textwidth]{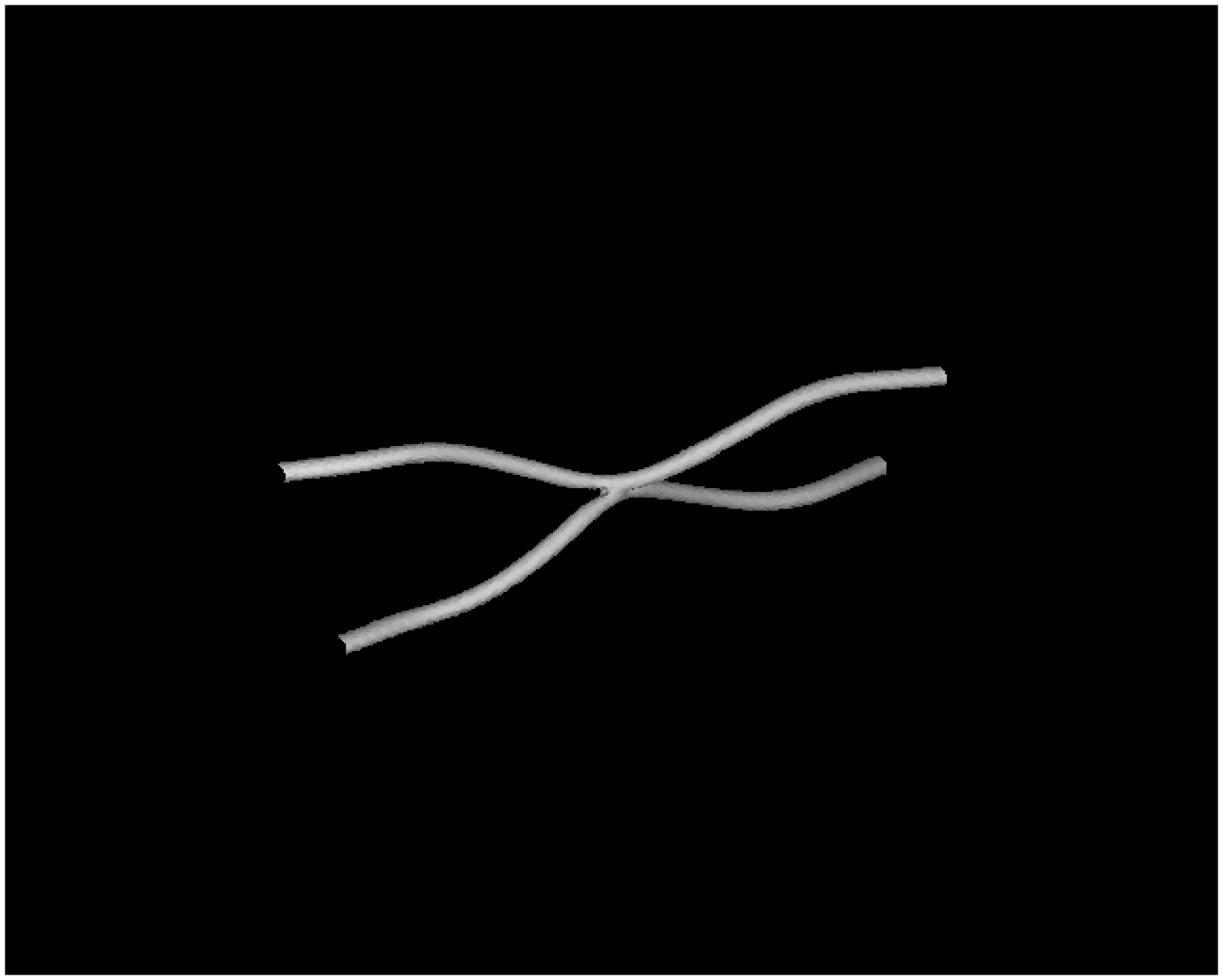}
\includegraphics*[width=.4\textwidth]{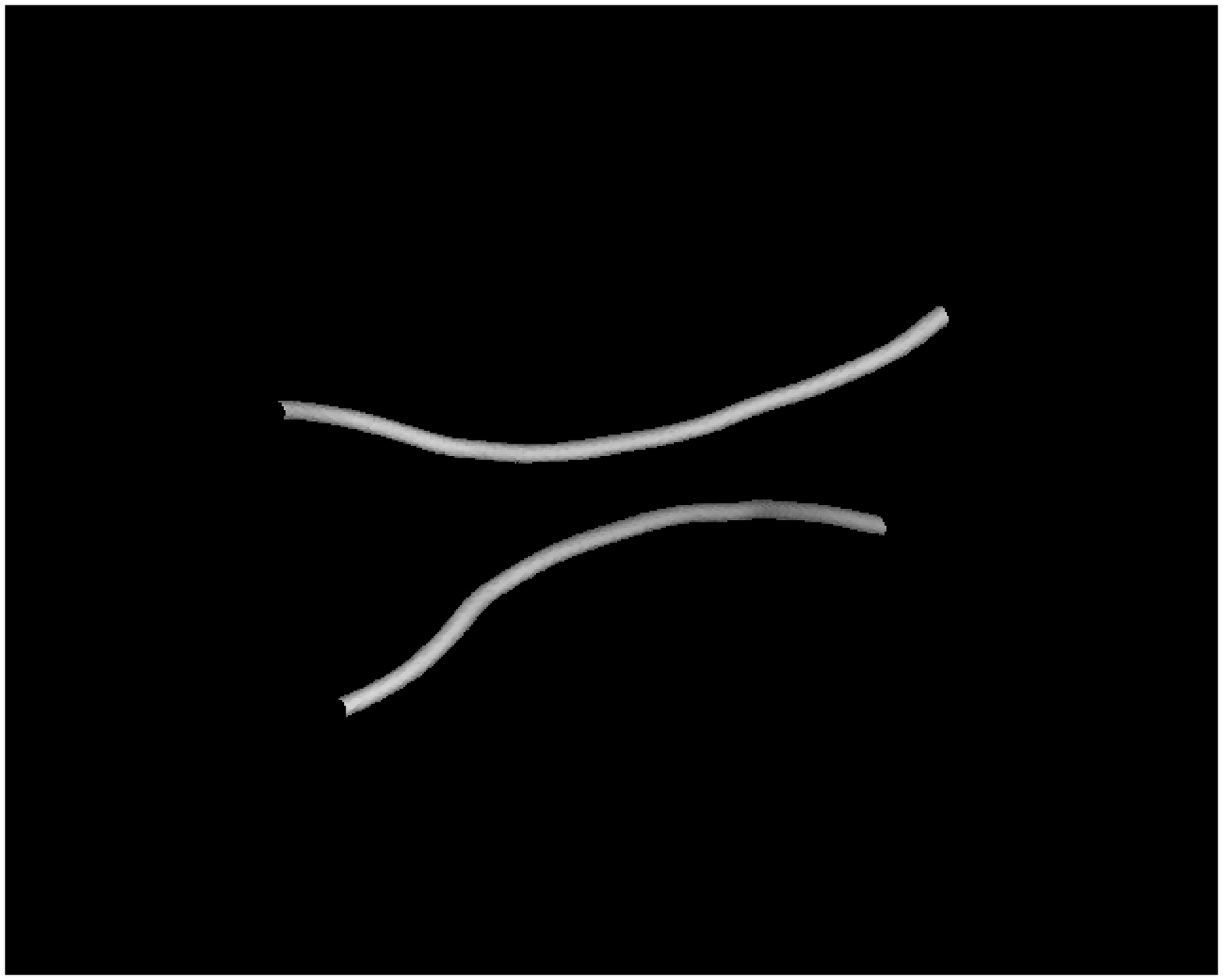}

\vspace{1.0cm}

\caption{\label{f:doubleintercommutation} Snapshots 
(from left to right and top to bottom) showing constant energy 
density isosurfaces in a simulation of two strings with 
winding $n=1$ at $\beta=0.125$ 
colliding with velocity $v=0.5$, at an angle $\alpha=10\,^{\circ}$, 
and undergoing double intercommutation, as indicated by 
the $\vartriangle$ in right hand Fig.~\ref{f:constraints}.}

\vspace{1.0cm}

\end{figure*}


\begin{figure*}
\centering

\includegraphics*[width=0.4\textwidth]{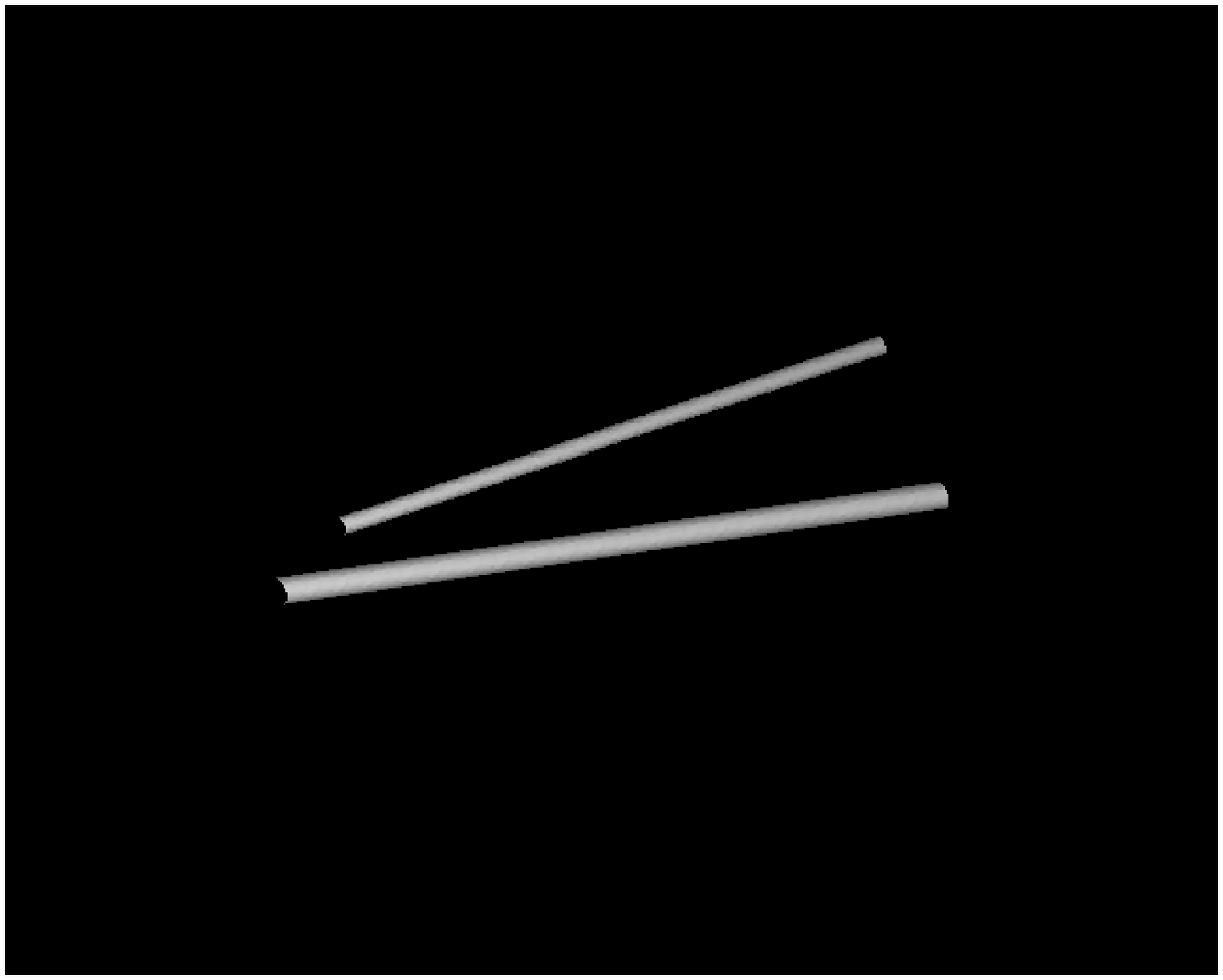}
\includegraphics*[width=0.4\textwidth]{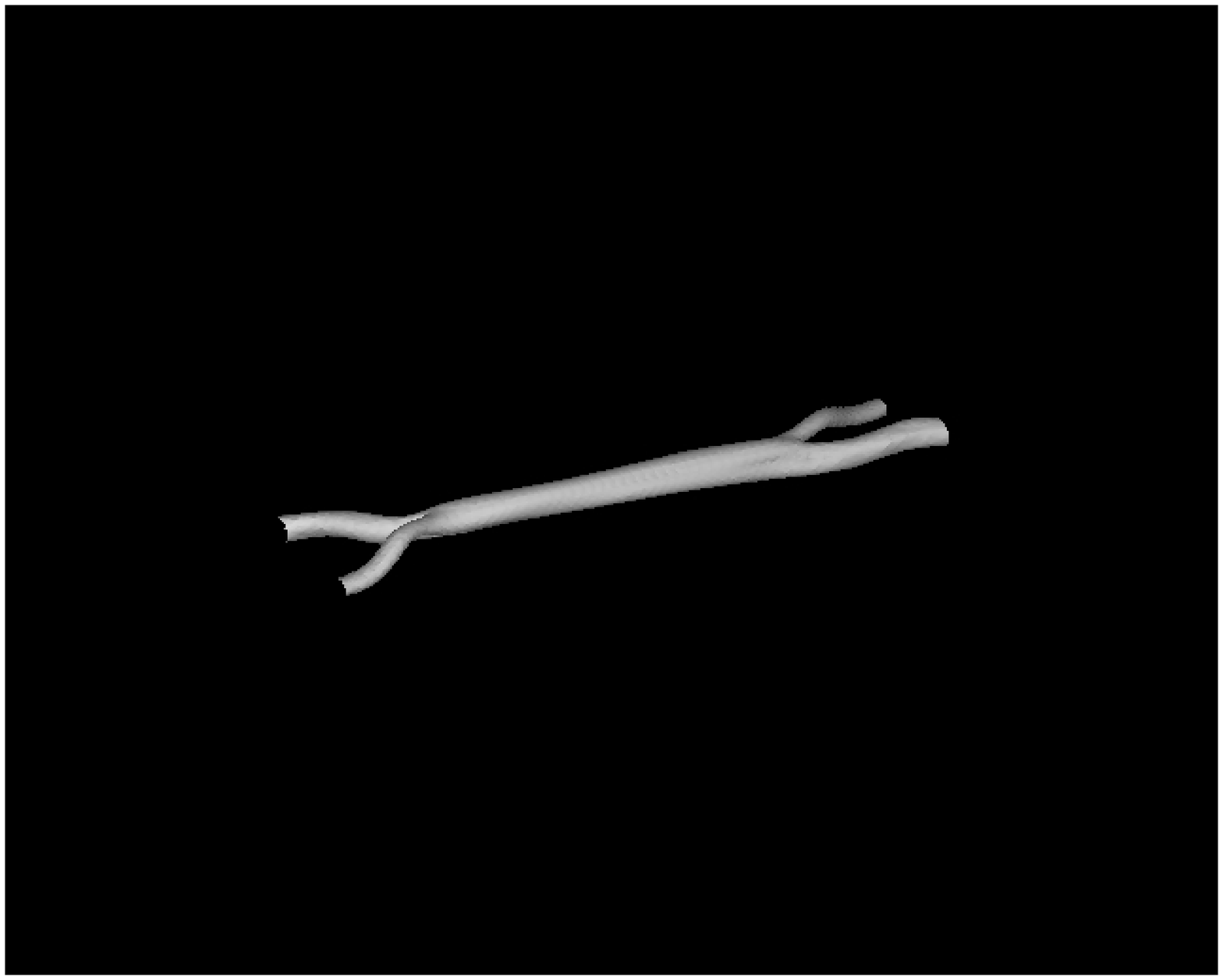}

\vspace{1.0cm}

\caption{\label{f:merge2} Snapshots (from left to right) showing 
constant energy density isosurfaces in a simulation of strings with winding 
$n=1$ (initially above) and $n=2$ at $\beta=0.125$ 
colliding with velocity $v=0.1$ and at an angle $\alpha=10 \,^{\circ}$.}

\end{figure*}


Local cosmic strings with an equal winding number, here $n=1$, 
always {\it intercommute} when intersecting. 
However, after intercommutation, strings can come together and merge to 
form a string of winding number $n=2$. 
This occurs when strings meet at relatively small 
velocity and angle (see~\cite{Bettencourt:1994kc,Bettencourt:1996qe}).
Alternatively strings only intercommute and never meet again.
In Figure~\ref{f:merge} four snapshots are presented that show the 
isosurfaces of constant energy density (set to be $0.2$ in dimensionless 
units, roughly 30\% of the maximum value in the core for a string with $n=1$, 
a value chosen so that strings with winding $n=1,2$ and $3$ can 
all be readily identified) when two strings with $n=1$ 
collide at velocity $v=0.1$ and angle $\alpha=15 \,^{\circ}$ 
and the coupling constant is $\beta=0.125$.

The formation of the bound state in the Abelian-Higgs model is not entirely a 
straightforward, instantaneous process, as already reported 
in~\cite{Bettencourt:1996qe}. 
The link forms and unforms, the two strings becoming separately visible again.
This is not surprising; the process is essentially right-angle scattering 
which is well-established by numerical experiments with 
vortices~\cite{Shellard:1988zx,Myers:1991yh}.
Analytically, right-angle scattering after a head-on collision 
is understood at critical coupling ($\beta=1$)~\cite{Ruback:1988ba}, 
in terms of geodesic motion in the moduli space approximation. 
This approach was introduced for the first time in the context of monopole 
scattering~\cite{Manton:1981mp} but has been recently 
re-employed in studies of strings in more complicated 
models~\cite{Hanany:2005bc,Eto:2006db}.
After intercommutation, here the string segments align almost parallel 
before coming together. Therefore locally the second collision takes place 
head-on with effectively zero impact parameter and strings scatter 
perpendicularly with respect to the incoming direction.
Repeated right-angle scatterings are clearly visible in the snapshots; 
the time-scale of attenuation is comparable to the time the simulations can 
be evolved.

The main result of this study is presented in Figure~\ref{f:constraints}, 
where the results from the simulations, standard intercommutation versus 
link-formation, are shown together with the prediction of~(\ref{inequality}) 
for the allowed parameter region in the $(\alpha,v)$-plane. 
A decrease in the scalar coupling $\lambda$ reduces 
the tension of a string 
with winding number $n=2$ relatively more than that of a string with $n=1$. 
This allows one to vary the ratio $\mu_3 / \mu_1$ to a certain extent 
by reducing the parameter $\beta$;
at $\beta=0.36$ we obtain $\mu_3 / \mu_1 \simeq 1.88$, whereas 
$\beta = 0.125$ yields $\mu_3 / \mu_1 \simeq 1.77$. This already 
leads to a considerable difference when the limiting curve 
of the equality in~(\ref{inequality}) is plotted for $\beta=0.36$ (left) 
and $\beta=0.125$ (right) in Figure~\ref{f:constraints}. 
Standard intercommutation events are indicated by crosses 
and the formation of links by circles. 
At $\beta=0.36$ the agreement between the simulations and 
the analytical prediction is perfect 
within the precision of the grid used in the $(\alpha,v)$-parameter space. 
There is a discrepancy at the largest angles when $\beta=0.125$; 
while the formation of an $x$-link should be allowed, this is not 
observed in the simulations.

This behaviour is to be expected from considerations of energy conservation. 
In the Nambu-Goto model, all the energy released by shortening of the 
colliding strings goes into the formation of the linking string. 
In the field-theory model, by contrast, some energy is radiated away. 
We therefore expect that it should be slightly harder to form a link, 
and so there should be a small band where link formation 
is possible for Nambu-Goto strings, but not for field-theory strings.

In addition, at $\beta=0.125$ there is a band just above the highest 
velocity allowing $x$-links to form suggested by~(\ref{inequality}) 
where a link does not form, but strings come together again 
and intercommute for a second time, events denoted by triangles 
in Figure~\ref{f:constraints}.
Snapshots of this process are presented in 
Figure~\ref{f:doubleintercommutation}.
The end result is thus two strings consisting of the same segments as 
initially and indistinguishable from the situation where strings had 
passed through each other apart from some deformation around 
the interaction point. 
A similar type of effective non-intercommutation has been reported 
when strings collide at very high 
velocities~\cite{Matzner:1988,Achucarro:2006es}. 
There is no sign of this kind of phenomenon at $\beta=0.36$. 
This may be because less energy is radiated away in the field 
when the scalar mass is larger.

The general case, where all three strings have different tensions can be 
investigated in the Abelian-Higgs model by colliding strings with different 
winding numbers. This was done for two strings with windings $n=2$ and $n=1$ 
at $\beta=0.125$.
Though the explicit analytic formula for the asymmetric 
case~\cite{Copeland:2007nv} is not presented here, 
the kinematically allowed area is almost degenerate with the one 
presented in Figure~\ref{f:constraints} for the coupling $\beta=0.125$. 
A complication in the Abelian-Higgs model is that due to unequal 
winding numbers, 
the intercommutation leads to a formation of a bridge between the strings 
(see also~\cite{Laguna:1989hn}), whose influence on bound-state 
formation at the very least cannot be entirely 
neglected.
No systematic study in the kinematic parameter space was carried out, but 
Figure~\ref{f:merge2} shows snapshots of a simulation that provides evidence 
for a bound state formation now in the form of a string with winding 
number $n=3$.

\section{Discussion}

We have reported on numerical simulations of string collisions in the 
Abelian-Higgs model. The objective was to monitor the kinematic 
parameter space $(\alpha,v)$ and compare the outcome to the analytical 
prediction. 
We do not observe $x$-links outside the area where they are not
expected to be kinematically allowed.
On the other hand, bound states generically form whenever allowed to appear. 
This is interesting because such links do not have to form dynamically; 
there could have been simple intercommutation events instead, but 
almost the whole region appears to prefer to form $x$-links.

It is not surprising that the observed discrepancy occurs when strings 
intersect at large angles - the Nambu-Goto action does not include the 
effects of intercommutation. As strings in the Abelian-Higgs model always 
intercommute, effectively the original strings break up, and when the new 
strings straighten after this reconnection, they do not come together 
and merge at large angles. 
This could be different for strings that do not intercommute and 
exchange partners and it would be interesting to see what a study 
with strings in two separate gauge fields would yield.

Secondly, we have demonstrated once more 
that the `effective' reconnection probability 
of strings even in the Abelian-Higgs model is not strictly 
unity. This is well-documented by numerical 
experiments at high collision speeds where a second 
intercommutation even takes 
place (see e.g.~\cite{Matzner:1988,Achucarro:2006es})
but as shown here can occur at moderate velocities too 
(it would be interesting to see if a modification of the 
moduli space approximation 
could capture the second 
intercommutation event). 
It was argued in~\cite{Bettencourt:1996qe} that when the 
bound state with winding $n=2$ dissolves 
(which inevitably happens eventually with the boundary conditions introduced 
in~\cite{Matzner:1988}), the original strings re-emerge. 
This is effectively like no intercommutation taking place.
However, the process reported here does not proceed via bound state formation 
and dissolution, but rather by repeated intercommutation, and seems 
therefore to be of a different nature.
This is further confirmed by the inequality~(\ref{inequality})
according to which the bound state is forbidden at velocities where the double 
intercommutation is observed.

Obviously the use of the Abelian-Higgs model has restricted us in this study 
to very limited values of the ratio of string tensions $ \mu_3 / \mu_1$, 
whereas models inspired by superstring theory would allow a much wider 
range of values.  We have shown that strings in the Abelian-Higgs model can 
have the same principal features as expected from cosmic superstrings, namely 
formation of junctions and effective reconnection probability 
not strictly unity even at moderately {\it low velocities}.
While network simulations very deep in the Type-I regime are hardly 
feasible to 
perform, these results may have implications for theories relying on a
very low value of the scalar coupling~$\lambda$. 
Such a network can arise e.g. upon breaking a gauge symmetry along a 
flat direction in supersymmetric theories
as has been pointed out recently in~\cite{Cui:2007js}.
To summarize, as anticipated already in~\cite{Bettencourt:1994kc}, this study 
demonstrates the richness of strings even in the simple Abelian-Higgs model.


\begin{acknowledgments}

P.~S. acknowledges Serena Bertone for the help in visualisation. 
A.~A., R.~P. and P.~S. were supported by 
the Netherlands Organization for Scientific Research (N.W.O.) 
under the VICI programme 
and A.~A. also partially by 
Spanish Ministry of Science and Technology project FPA2005-04823
and Basque Government project IT-357-07.
The simulations have been performed at the Lisa cluster, 
its use permitted by an NCF grant. 

\end{acknowledgments}


\bibliography{references}
\include{references}

\end{document}